\documentclass[fleqn,12pt,twoside]{article}
\usepackage{espcrc1}

\usepackage{graphicx}
\usepackage[figuresright]{rotating}

\newcommand{\AmS}{{\protect\the\textfont2
  A\kern-.1667em\lower.5ex\hbox{M}\kern-.125emS}}
\newcommand{\solar}{\ensuremath{_{\odot}}}

\title{The Innermost Ejecta of Core Collapse Supernovae}

\author{C.\ Fr\"ohlich\address[unibas]{Dept.\ of Physics and Astronomy, 
        Univ.\ of Basel,
        Basel, Switzerland}%
        \thanks{Support by the Swiss NSF grant 2000-061031.02
                },
        P.\ Hauser\addressmark ,
        M.\ Liebend\"orfer\address{CITA,
        University of Toronto,
        Toronto ON, Canada},
        G.\ Mart\'{\i}nez-Pinedo\address[barcelona]{ICREA and Institut d'Estudis Espacials de Catalunya, Barcelona, Spain},
        E.\ Bravo \address{Universitat Polit\`ecnica de Catalunya, Barcelona, Spain},
        W.R.\ Hix\address{Physics Division, Oak Ridge National Laboratory,
        Oak Ridge TN, USA},
        N.T.\ Zinner \address{Institute of Physics and Astronomy,
        Aarhus University, Aarhus C, Denmark}
        and
        F.-K.\ Thielemann\addressmark[unibas]
        }
       
\begin{document}

\maketitle

\begin{abstract}
We ensure successful explosions (of otherwise non-explosive models) by enhancing the neutrino luminosity via reducing the neutrino scattering cross sections or by increasing the heating efficiency via enhancing the neutrino absorption cross sections in the heating region.
Our investigations show that the resulting electron fraction $Y_e$ in the innermost ejecta is close to 0.5, in some areas even exceeding 0.5.
We present the effects of the resulting values for $Y_e$ on the nucleosynthesis yields of the innermost zones of core collapse supernovae.

\end{abstract}

\section{Introduction}

Presently, supernova explosions cannot be explained in 1D spherically \cite{Ramp00,Mezz01,Lieb01} or 2D rotationally \cite{Mezz98,Bura03} symmetric radiation-hydrodynamics calculations.
Besides the need for further refinement and consolidation of the current simulations, successful explosion simulations may rely on rotation and magnetic fields \cite{Thom00,Thom04} or unknown microphysics, e.g. for the neutrino interactions.
For the prediction of the nucleosynthesis yields in supernovae, we face the problem how to simulate the behavior of ejected matter realistically, given the existing problems to obtain explosions in self-consistent models with accurate neutrino physics.
Artificially induced explosions for supernova nucleosynthesis predictions \cite{Thie90,Thie96,Woos95} are a valid approach for the outer layers,
but are inconsistent for the innermost layers, affecting the Fe-group composition.
The nucleosynthesis in these zones (undergoing Si-burning) is dominated by the electron fraction $Y_e$.
The constraints from observations require the resulting $Y_e$ in the innermost zones to be $\ge 0.5$ \cite{Thie96}.


\begin{figure}[t]
\begin{center}
\includegraphics*[width=15cm]{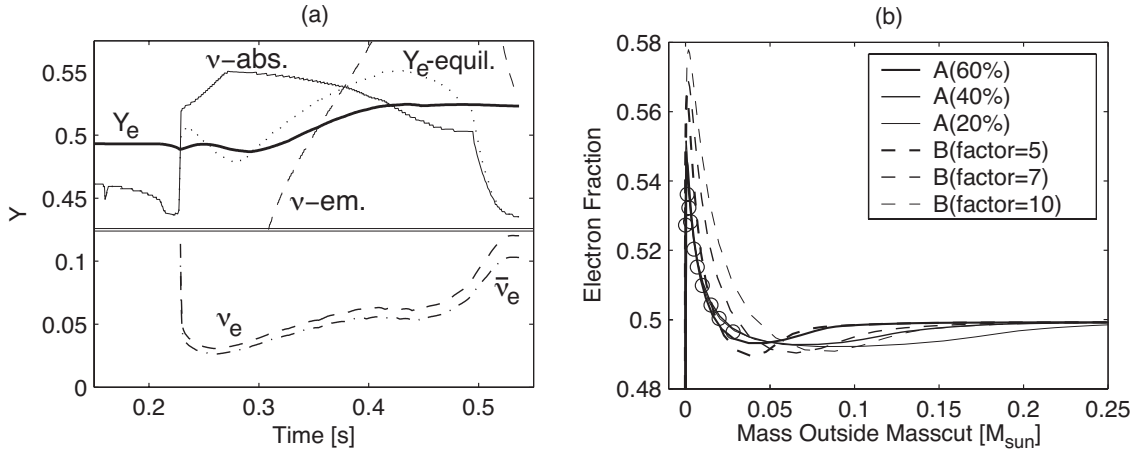}
\caption{{\bf (a)}
The upper panel shows the time evolution of the electron fraction in a fluid element close to the mass cut (thick solid line).
The dotted line is the equilibrium electron fraction for infinitely long exposure to the neutrino field.
The dashed and thin solid lines represent equilibrium with only neutrino absorption or emission considered.
The electron fraction is determined by competition between the neutrino interaction rates and the matter ejection timescale.
{\bf (b)}
Electron fraction as function of mass when the runs are stopped (0.4s -- 0.6s after bounce) with
(A) reduction of the neutral current scattering opacities by a given percentage (solid lines), or
(B) enhancement of the reaction timescales (dashed lines).
The open circles are final electron fractions for model A(40\%).
}
\label{fig-trajectory}
\end{center}
\end{figure}

\begin{figure}[t]
\begin{center}
\includegraphics*[height=12cm,angle=-90]{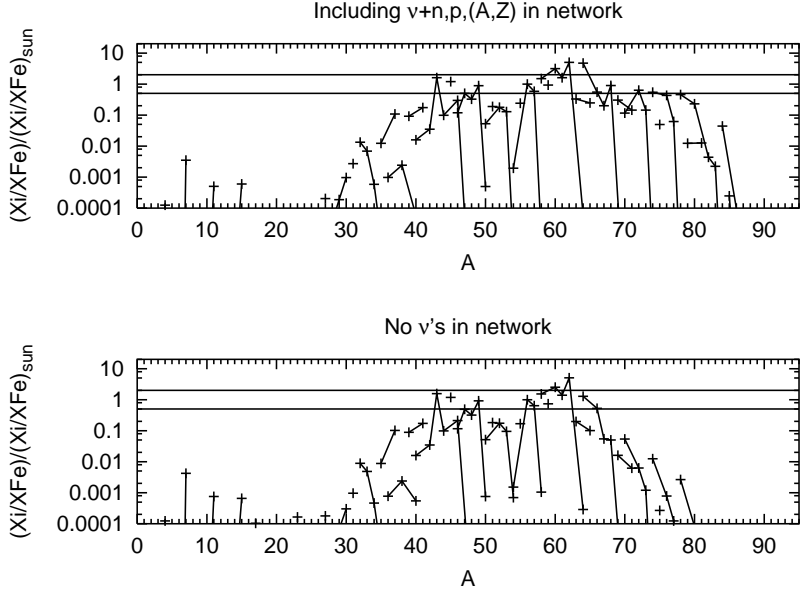}
\caption{Mass fractions of the stable nuclei for model A(40\%) normalized to solar abundances.
For both calculations shown the postprocessing is based on the same hydro profile.
The differences originate from the set of reactions included in the nuclear reaction network:
no neutrino-induced reactions included, or neutrino captures on free nucleons and on nuclei included.}
\label{fig-nucleo}
\end{center}
\end{figure}


\section{Hydrodynamical Simulations}

The simulations are performed in spherical symmetry with general relativistic Boltzmann neutrino transport using the code AGILE-BOLTZTRAN.
For a detailed description of the code refer to \cite{Lieb04}.
We use two different approaches to simulate explosions.
In the first approach, the neutral current scattering opacities of neutrinos on free nucleons are scaled with a factor ranging from 0.1 to 0.7 while using standard cross sections for electron/positron/neutrino/anti-neutrino capture on free nucleons (series A
\footnote{Note that series A has been calculated at reduced neutrino angle resolution which additionally enhanced the neutrino abundances.}).
This leads to a faster deleptonization of the protoneutron star such that the neutrino luminosities are boosted in the heating region,
as it might happen with rigorous protoneutron star convection or dramatic changes in the input physics at high densities.
In the second approach, the charged current neutrino reactions (and their inverse reactions) are increased by equal factors in the heating region (series B) to crudly mimic a prolonged exposure to neutrino heating in the convective heating region which cannot be appropriately treated in spherically symmetric simulations.
This enhances the time scale for neutrino heating without changing the equilibrium $Y_e$.

The models are based on a progenitor of $20 \mathrm{M\solar}$ \cite{Nomo88}.
For each series the choice of parameters is such that they correspond to a barely exploding model, an extremely exploding model, and a model with average explosion properties.
The explosion energies are of the order of $10^{51} \mathrm{ergs}$.
After $\sim 100 \mathrm{ms}$ the accretion front stalls in all models and the explosion is launched after a delay of additional $\sim 100 \mathrm{ms}$.
Three different phases can be distinguished in the mass trajectory:
During the first phase the mass element falls inwards in the gravitational potential.
This phase lasts $\sim 200 \mathrm{ms}$ and ends with an almost instantaneous deceleration as it falls through the accretion shock.
For approximately $200 \mathrm{ms}$ it drifts around in the heating region until, in a third phase, it is ejected to larger radii lifting the electron degeneracy.
When the electron chemical potential falls below the mass difference between neutrons and protons, protons are favored over neutrons \cite{Belo03}.
Figure \ref{fig-trajectory} compares the evolution of the electron fraction to equilibrium values obtained under different assumptions.
The simulations are stopped when convergence problems appear due to the mass separation between the remnant and the ejecta.

\section{Nucleosynthesis}

For the nucleosynthesis results presented here, we consider only the first few zones outside of the mass cut.
The position of the mass cut emerges consistently from the simulation as the region of bifurcation in which the density has dropped below $\sim 10^6 \mathrm{g/cm^3}$.
Based on the hydrodynamics profiles, the detailed nucleosynthesis is calculated in a framework of consistently treating all weak interactions, namely neutrino/anti-neutrino absorption on free nucleons, the inverse reactions thereof [electron/positron capture], and neutrino/anti-neutrino absorption on nuclei.

All of the models develop a proton-rich region around the mass cut.
This can be seen in Figure \ref{fig-trajectory}(b) which shows the electron fraction $\sim 0.5\mathrm{s}$ after bounce.
The open circles denote the final electron fraction of the exemplary run A(40\%) at temperatures $T<0.1 \times 10^9 \mathrm{K}$.

Figure \ref{fig-nucleo} shows the mass fractions of the stable nuclei for model A(40\%).
In this model, the mass cut sits at $1.511 \mathrm{M\solar}$.
For the other models that mass cut is between $1.44 \mathrm{M\solar}$ and $1.586 \mathrm{M\solar}$.
The light elements are mainly produced in the outer zones.
Here, we are only concerned with the Fe-group composition.
For the intermediate mass elements the main improvement compared to earlier calculations is the higher production of single nuclei like $\mathrm{^{45}Sc}$ and $\mathrm{^{51}V}$.
Neutrino induced reactions become important for the nucleosynthesis of nuclei in the mass range $A>70$.
These nuclei are mainly produced in the zones close to the mass cut where the electron fraction strongly depends on the neutrino captures.

\section{Conclusions}
We have ensured explosions (of otherwise non-explosive models) by changing the neutrino luminosity in two different ways.
Detailed nucleosynthesis calculations based on a consistent treatment of all weak interactions show an electron fraction $Y_e\sim0.5$ for the innermost ejected zones as it is consistent with observations.
Our calculations confirm the importance of consistent treatment of neutrino induced interactions for the nucleosynthesis of the innermost zones.
In the Fe-group improvement was achieved for the isotopes $^{58,61,62}\mathrm{Ni}$, $^{51}\mathrm{V}$, and $^{45}\mathrm{Sc}$.
However, the nucleosynthesis calculations based on the innermost $0.03 \mathrm{M\solar}$ remain to be extended to the outer layers to obtain a complete prediciton.


\begin{thebibliography}{9}
\bibitem{Ramp00} M.\ Rampp and H.-T.\ Janka
                 ApJ 539 (2000) L33
\bibitem{Mezz01} A.\ Mezzacappa, M.\ Liebend\"orfer, O.E.B.\ Messer,
                 W.R.\ Hix, F.-K.\ Thielemann and S.W.\ Bruenn,
                 Phys. Rev. Lett. 86 (2001) 1935
\bibitem{Lieb01} M.\ Liebend\"orfer, A.\ Mezzacappa, F.-K.\ Thielemann,
                 O.E.B.\ Messer, W.R.\ Hix, S.W.\ Bruenn,
                 Phys. Rev. D 63 (2001) 103004
\bibitem{Mezz98} A.\ Mezzacappa, A.C.\ Calder, S.W.\ Bruenn, J.M.\ Blondin,
                 M.W.\ Guidry, M.R.\ Strayer and A.S.\ Umar,
                 ApJ 495 (1998) 911
\bibitem{Bura03} R.\ Buras, M.\ Rampp, H.-Th.\ Janka and K.\ Kifonidis,
                 Phys. Rev. Lett. 90 (2003) 1101
\bibitem{Thom00} C.\ Thompson
                 ApJ 534 (2000) 915
\bibitem{Thom04} T.A.\ Thompson, E.\ Quataert and A.\ Burrows,
                 astro-ph/0403224
\bibitem{Thie90} F.-K.\ Thielemann, M.\ Hashimoto and K.\ Nomoto,
                 ApJ 349 (1990) 222
\bibitem{Thie96} F.-K.\ Thielemann, K.\ Nomoto and M.\ Hashimoto,
                 ApJ 460 (1996) 408
\bibitem{Woos95} S.E.\ Woosley and T.A.\ Weaver,
                 ApJS 101 (1995) 181
\bibitem{Lieb04} M.\ Liebend\"orfer, O.E.B.\ Messer, A.\ Mezzacappa,
                 S.W.\ Bruenn, C.Y.\ Cardall and F.-K.\ Thielemann,
                 ApJS 150 (2004) 263
\bibitem{Nomo88} K.\ Nomoto and M.\ Hashimoto,
                 Phys. Rep. 163 (1988) 13
\bibitem{Belo03} A. Beloborodov,
                 ApJ 588 (2003) 931
\end{thebibliography}
\end{document}